\documentclass[3p,times,procedia]{elsarticle}
\usepackage{nupha_ecrc}

\volume{00}

\firstpage{1}

\journalname{Nuclear Physics A}

\runauth{}

\jid{nupha}

\jnltitlelogo{Nuclear Physics A}

\usepackage{amssymb}

\usepackage[figuresright]{rotating}

\begin{document}

\begin{frontmatter}



\dochead{}

\title{$p\Xi^- $ Correlation in Relativistic Heavy Ion Collisions\\
with Nucleon-Hyperon Interaction from Lattice QCD
 }


\author{Tetsuo Hatsuda$^1$}
\author{Kenji Morita$^2$}
\author{Akira Ohnishi$^2$}
\author{Kenji Sasaki$^2$}

\address{$^1$ iTHEMS Program and Nishina Center, RIKEN, Saitama 351-0198, Japan }
\address{$^2$ Yukawa Institute for Theoretical Physics, Kyoto University,
Kyoto 606-8502, Japan}

\begin{abstract}
On the basis of the   $p\Xi^-$ interaction extracted from 
 (2+1)-flavor lattice QCD simulations at the physical point, 
  the momentum  correlation of $p$ and $\Xi^-$  produced in 
  relativistic heavy ion collisions is evaluated. 
    $C_{\rm SL}(Q)$  defined by a ratio of the momentum correlations
     between the systems with different source sizes  is shown to be largely enhanced
     at low momentum 
 due to the strong attraction between $p$ and $\Xi^-$ in the
    $I=J=0$ channel.  Thus, measuring this ratio  at RHIC and LHC and its comparison to the
 theoretical analysis will give a useful constraint 
 on the $p\Xi^-$ interaction. 
 \end{abstract}


\end{frontmatter}


\section{Introduction}
\label{}

The coupled-channel
  Nambu-Bethe-Salpeter (NBS) wave function measured in lattice QCD \cite{Ishii:2006ec,HALQCD:2012aa}
can now provide ``theoretical" information of hyperon-nucleon and hyperon-hyperon
 interactions  through the HAL QCD method \cite{Doi:2017cfx,Sasaki:2017ysy,Ishii:2017xud,Nemura:2017bbw}. 
The energy-independent non-local potentials $U(r, r')$  obtained by the method allow us to
  calculate the scattering phase shifts and binding energies of two baryons.
 
 These potentials are also useful for  analyzing the two-particle momentum correlations 
  in relativistic heavy ion collisions \cite{Cho:2017dcy}. 
 It was recently  studied in \cite{Morita:2016auo} that the possible spin-2 $p\Omega^-$ dibaryon state 
 suggested by lattice QCD  \cite{Etminan:2014tya} can be probed
  by the $p\Omega^- $ momentum correlation at RHIC and LHC.
    In particular, the ratio of correlation functions between small and large collision systems,
$C_{\rm SL}(Q)$, is shown to be  a  good measure to extract the strong interaction effect 
without much contamination from the Coulomb effect  \cite{Morita:2016auo}.
In the present paper, we extend the analysis to the  $p\Xi^-$ system in
$I=J=0$ channel which was recently predicted to have large attraction by the lattice QCD simulations
 at physical quark masses \cite{Sasaki:2017ysy}.
 
\section{Lattice QCD formulation}

We start with the normalized four-point function $R$ in channel $\alpha$ defined by
\begin{eqnarray}
 R^{\alpha}(\vec r, t) 
 &\equiv& \frac{\langle 0 \mid B_{\alpha_1}(\vec{x} +\vec{r},t) B_{\alpha_2}(\vec{x}, t) \ \overline{\mathcal{J}} (0)\mid 0 \rangle}{ \sqrt{Z_{\alpha_1} Z_{\alpha_2}} \exp[-(m_{\alpha_1}+m_{\alpha_2}) t]}, 
\label{EQ.Rdefine}
\end{eqnarray}
where $B_{\alpha_1}(\vec{x},t)$  and $B_{\alpha_2}(\vec{x},t)$  
are the sink operators for octet baryons. 
  $\sqrt{Z_{\alpha_1}}$ $\sqrt{Z_{\alpha_2}}$ 
  are the corresponding wave-function renormalization factors, and
${\mathcal{J}} (0)$ is a source operator at zero  initial-time  to create two baryons. 
 The coupled channel potential is obtained through the linear partial differential equation ~\cite{HALQCD:2012aa};
\begin{eqnarray}
\left(
D_t^\alpha - H_0^\alpha \right) R^{\alpha}(\vec r, t)  &=& \int d^3 r^\prime \ U^{\alpha \beta}(\vec r, \vec r^\prime) \Delta^{\alpha \beta}
R^{\beta}(\vec r^\prime, t),
\end{eqnarray}
with $ {H_0}^{\alpha} = - \frac{\nabla^2}{2\mu^{\alpha}} $ and $\Delta^{\alpha \beta}
 = \exp[-(m_{\beta_1}+m_{\beta_2})t] /  \exp[-(m_{\alpha_1}+m_{\alpha_2})t] $. 
   $D_t^{\alpha}$ is a time-derivative operator whose leading-order term
 reads  $ -\partial /\partial t$.
 We introduce a derivative expansion to treat the non-local potential as
\begin{eqnarray}
  U^{\alpha \beta}(\vec r,\vec r') 
  = ( V_{\rm LO}^{\alpha \beta} (\vec r)+ V_{\rm NLO}^{\alpha \beta}(\vec r)  + \cdots )\delta(\vec r-\vec r^\prime). 
\end{eqnarray}
In the following, we truncate the expansion at the leading order.

We employ $(2+1)$-flavor QCD configurations 
 on the $L^4 = 96^4 $ lattice with the lattice spacing $a \simeq 0.085$fm.
 This corresponds to the physical size, $La = 8.1{\rm{fm}}$, which 
 guarantees that  the finite volume effect   on $U^{\alpha \beta}(\vec r,\vec r')$  is negligible. 
   The quark masses are chosen for the system 
 to be   almost at the physical point; 
 $m_\pi \simeq 146$ MeV and $m_K \simeq 525$ MeV \cite{Sasaki:2017ysy}. 
The total number of configurations is $414$ $\times$ $4$ space-time rotations $\times$ $48$ wall sources.
The baryon masses measured in this setup are listed below.

\begin{center}
\begin{tabular}{c|cccc}
\hline \hline
baryon & $N$ & $\Lambda$ & $\Sigma$ & $\Xi$ \\
\hline
mass [MeV] & 953 $\pm$ 7 & 1123 $\pm$ 3 & 1204 $\pm$ 2 & 1332 $\pm$ 1 \\
\hline \hline
\end{tabular}
\end{center}
\label{tab:mass}

\section{$p\Xi^-$ potential in $I=0$ channel}

\begin{figure}[t]
\begin{center}
\includegraphics[scale=0.4]{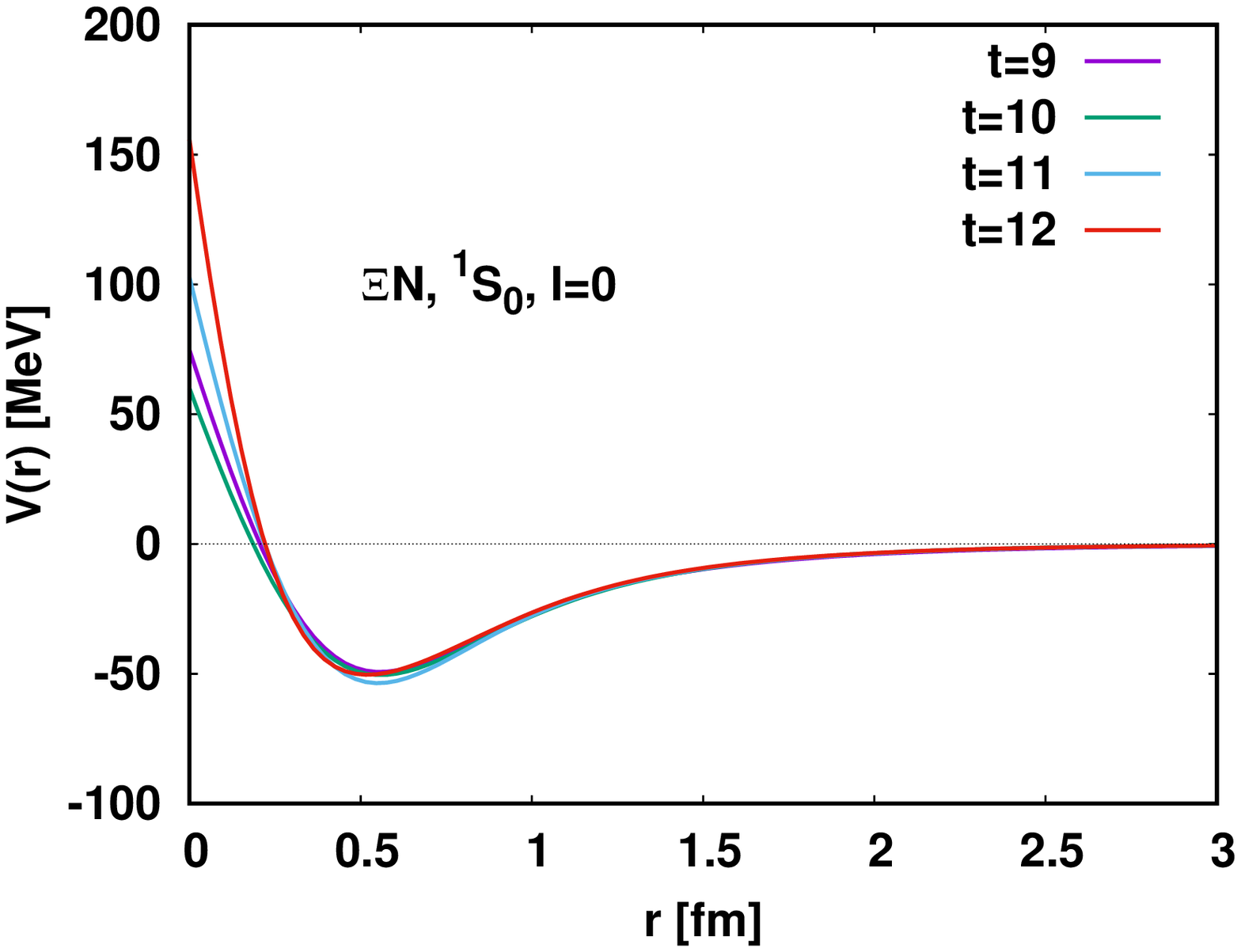}
\includegraphics[scale=0.4]{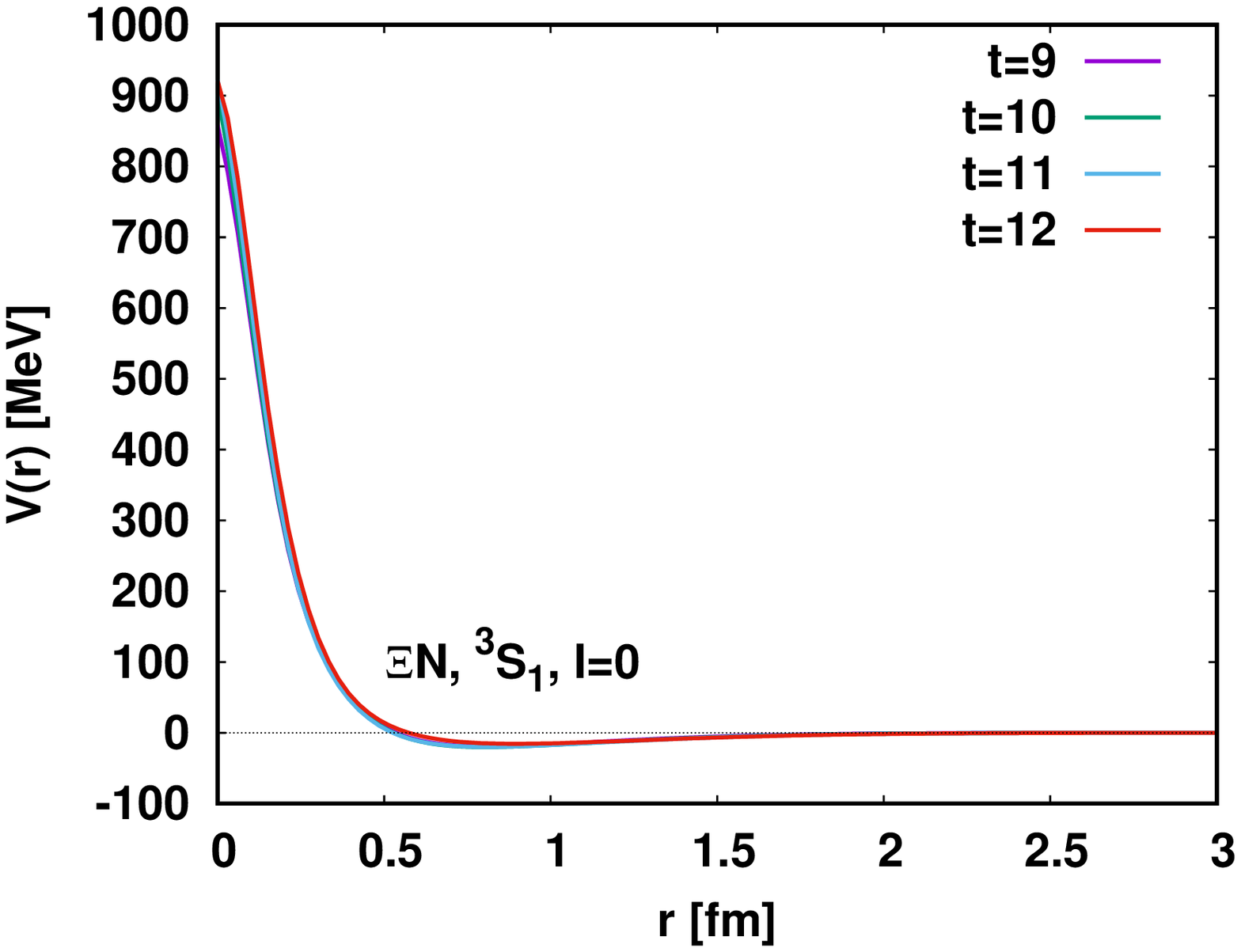}
\vspace{0.3cm}
\caption{The $N\Xi$ potentials in the $I=0$ channel fitted to the (2+1)-flavor lattice QCD data at the 
physical point.   Euclidean time used for extracting the lattice QCD potential is denoted by $t$.
(Left) The potential in the $(I,J)=(0,0)$ channel ($^1 S_0$). 
(Right) The potential in the $(I,J)=(0,1)$ channel ($^3 S_1$). 
 }
\label{fig:potentials}
\end{center}
\end{figure}

The $S=-2$ baryon-baryon interactions including the I=0 $\Lambda\Lambda-N\Xi-\Sigma\Sigma$ coupled-channel
 system have been recently reported in \cite{Sasaki:2017ysy}.  
In particular, one of the diagonal components 
$V_{N\Xi,N\Xi}(r)$ in the $(I,J)=(0,0)$  channel  ($^1 S_0$) was shown to have  large attractive well at intermediate
distance and relatively weak repulsive core at short distance, while
$V_{N\Xi,N\Xi}(r)$ in the $(I,J)=(0,1)$ channel ($^3 S_1$)   has weaker attractive well and stronger repulsive core.
 Also, $V_{N\Xi,N\Xi}(r)$ in the $I=1$ channels do not have appreciable attraction.
Motivated by these observations,
  we parametrize the lattice results of $V_{N\Xi,N\Xi}(r)$ in the $I=0$ channels by a
  combination of the Gauss and Yukawa functions as shown in Fig.\ref{fig:potentials}.  
 Curves with different $t$ correspond to the potentials obtained from $R({\vec x},t)$ for
 different $t$, so that the $t$ dependence of $V(r)$ reflects typical magnitude of the systematic error
  of the lattice data. 
  We found that the strong QCD attraction in Fig.\ref{fig:potentials}(Left) together with the Coulomb attraction
 leads to the $^1 S_0$ system close to the unitary region where the inverse of the scattering length is
 close to zero. On the other hand,  the $^3S_1$  system described by  Fig.\ref{fig:potentials}(Right) has 
 strong repulsion even with the Coulomb attraction.

\section{$p\Xi^-$ momentum correlation}

The correlation function of non-identical pair such as $p\Xi^-$ 
 is given in terms of the two-particle distribution
$N_{p\Xi}({\bf k}_p,{\bf k}_\Xi)$ normalized by a 
product of the single particle distributions,
$N_\Xi({\bf k}_\Xi) N_p({\bf k}_p)$, 
 \begin{eqnarray}
 C({\bf Q},{\bf K}) \equiv
  \frac{N_{p\Xi}({\bf k}_p,{\bf k}_\Xi)}{N_p({\bf k}_p)
  N_\Xi({\bf k}_\Xi)}  \label{eq:full-CQ}
  \simeq \frac{ \int d^4x_p \int d^4x_\Xi \ S_p(x_p, {\bf k}_p)
  S_\Xi(x_\Xi, {\bf k}_\Xi) 
  \left| \Psi_{p\Xi}({\bf r}^\prime)\right|^2}
  {\int d^4x_p \ S_p(x_p,{\bf k}_p) \int d^4x_\Xi \ S_\Xi(x_\Xi, {\bf k}_\Xi)},
 \nonumber
\end{eqnarray}
where relative and total momenta are defined as
$ {\bf Q} = (m_p {\bf k}_\Xi - m_\Xi {\bf k}_p)/M$
 and  $ {\bf K} = {\bf k}_p + {\bf k}_\Xi$, respectively,  
with $M\equiv m_p + m_\Xi$.
The source functions
$S_i(x_i, {\bf k}_i)\equiv E_i \frac{dN_i}{d^3{\bf k}_i d^4 x_i} $ (with $i=p, \Xi$
 and $E_i= \sqrt{{\bf k}_i^2+m_i^2}$) correspond to the phase space distributions
of $p$ and $\Xi$ at freeze-out.  The final state interaction
after the freeze-out is described by the two-particle wave
function $\Psi_{p\Xi}$ with  a shifted relative coordinate
${\bf r}^\prime = {\bf x}_\Xi-{\bf x}_p-{\bf K}(t_p-t_\Xi)/M$.

\begin{figure}[t]
\begin{center}
\includegraphics[scale=0.45]{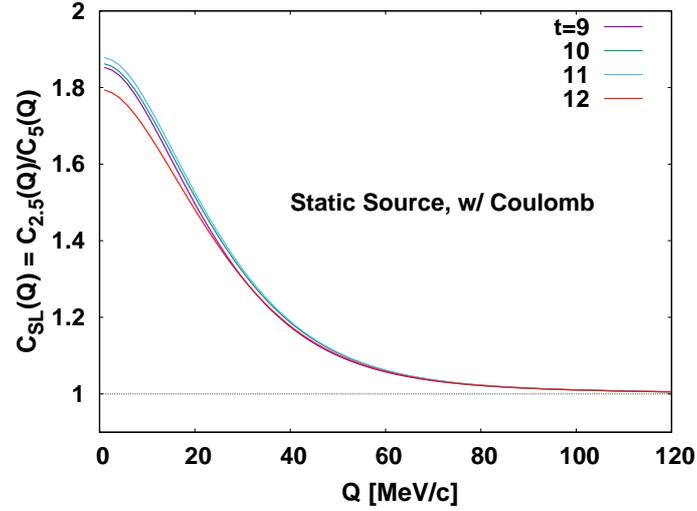}
\vspace{0.3cm}
\caption{SL (small-to-large) ratio $C_{\rm SL}(Q)$ for the momentum correlation of $p\Xi^-$ system
 as a function of the relative momentum $Q$ in the case of the static source.
 Both the strong and Coulomb interactions are taken into account for the $p\Xi^-$ interaction.
 Different curves correspond to different potentials shown in Fig.\ref{fig:potentials}.
}
\label{fig:CSL}
\end{center}
\end{figure}

Here we consider the static source function with spherical symmetry
to extract the essential part of physics;
\begin{equation}
 S_i(x_i,{\bf k}_i) 
 \propto  E_i \ e^{-\frac{{\bf x}_i^2}{2R_i^2}} \ \delta(t-t_i), \ \ (i=p, \Xi^-),
 \label{eq:staticsource}
\end{equation}
where $R_i$ is a source size parameter.
Assuming the equal-time emission $t_p=t_\Xi$, 
we obtain
 \begin{eqnarray}
C(Q) & =&  \int [dr] \int \frac{d\Omega}{4\pi}  |\psi^C({\bf r})|^2 
   + \frac{1}{8}  \int [dr]
   \left( |\chi_{\rm sc}^{\rm J=0}(r)|^2 - |\psi_{0}^C(r)|^2 \right)
 + \frac{3}{8} \int [dr]
  \left( |\chi_{\rm sc}^{\rm J=1}(r)|^2 - |\psi_{0}^C(r)|^2 \right) ,
 \label{eq:c2static}
\end{eqnarray}
where $[dr]=\frac{1}{2\sqrt{\pi}R^3} \! dr \, r^2  e^{-\frac{r^2}{4R^2}}$
 with $R = \sqrt{(R_p^2 + R_\Xi^2)/2}$ being the effective size
parameter. $\int d\Omega$ is the integration over the solid angle between
${\bf Q}$ and ${\bf r}$.  
Note that $\psi^C({\bf r})$ is the Coulomb wave function
 characterized by the reduced mass  and the Bohr radius of the $p\Xi^-$ system.
Its S-wave component is denoted by $\psi_{0}^C(r)$.
The scattering wave functions obtained by solving
the Schr\"{o}dinger equation with both  strong interaction  and  Coulomb interaction
 are denoted by   $\chi_{\rm sc}^{J=0}(r)$ and $\chi_{\rm sc}^{J=1}(r)$ for the $^1 S_0$ channel
 and $^3 S_1$ channel, respectively.
  We assume that
  the $I=1$ sector does not contribute substantially to $C(Q)$, which is
 supported  by the fact that the $I=1$ $p\Xi^-$ potential  has only short-range repulsion  \cite{Sasaki:2017ysy}.
 The factors $1/8=1/2 \times 1/4$ and $3/8=1/2 \times 3/4$ originate  from the isospin and spin multiplicities.
 Also, we assume that the absorptive contribution by the coupling to the  $\Lambda\Lambda$ channel
 is negligible since it is reported to be weak due to its short range nature  \cite{Sasaki:2017ysy}.
 
 In \cite{Morita:2016auo}, the ``SL (small-to-large) ratio'' was 
introduced:  It is defined as a ratio of $C(Q)$ 
between the systems with different source sizes,
\begin{eqnarray}
C_{\rm SL}(Q)\equiv  C_{R_{p,\Xi}=2.5{\rm fm}}(Q) / C_{R_{p,\Xi}=5{\rm fm}}(Q),
\end{eqnarray}
which has  good 
sensitivity to the strong interaction without much contamination from the  Coulomb interaction \cite{Morita:2016auo}. 
Shown in Fig.\ref{fig:CSL} is $C_{\rm SL}(Q)$ of the $p\Xi^-$ system with the Coulomb interaction under the
assumption of the static source given in  Eq.(\ref{eq:staticsource}).

The large enhancement of this ratio at small $Q$ originates from the fact that the $p\Xi^-$ system in the $^1 S_0$ 
channel is close to the unitary region. The result has rather weak dependence on $t$, which indicates that 
 the systematic errors of the lattice data do no aftect the  final results
  significantly.  We have also checked that taking the expanding source
as discussed in \cite{Morita:2016auo} does not change the present result.
 
\section{Summary}
The  momentum correlation of the $p\Xi^-$ system was presented 
 by employing the  $p\Xi^-$ potential extracted from the coupled channel analysis of the 
  (2+1)-flavor lattice QCD data at the physical point.  So-called the SL-ratio of the momentum 
  correlation ($C_{\rm SL}(Q)$) was calculated and was shown to have large enhancement 
   at small $Q$ due to the strong attraction between $p$ and $\Xi^-$ in the
    $^1 S_0$ channel.  Measuring this ratio  at RHIC and LHC and its comparison to the present
 theoretical analysis will give useful constraint 
 on the $p\Xi^-$ interaction.  Such information is particularly important not only for the
  nature of the possible $H$-dibaryon coupled to  $p\Xi^-$  \cite{Sasaki:2017ysy}
  but also for the properties of
   $\Xi$-hypernuclei \cite{Nakazawa:2015joa} 
   and  for   $\Xi^-$ in the central core of the neutron star \cite{Baym}.

\section*{Acknowledgments}
This work is supported in part by MEXT Grant-in-Aid for Scientific Research (JP15K17667, 24105008), 
SPIRE (Strategic Program
for Innovative REsearch) Field 5 project and "Priority Issue on Post-K computer" (Elucidation of the Fundamental Laws and Evolution of the Universe). K.M. was supported by JSPS Grant 16K05349, and 
National Science Center, Poland under grants:, Maestro DEC-2013/10/A/ST2/00106.
T.H. was partially supported by RIKEN iTHES Project and iTHEMS Program.
The authors thank HAL QCD Collaboration for proving us with the data for $p\Xi^-$ interactions and for stimulating discussions.

\bibliographystyle{elsarticle-num}

\end{document}